\documentstyle[twoside,fleqn,espcrc2]{article}

\newcommand{\AmS}{{\protect\the\textfont2
  A\kern-.1667em\lower.5ex\hbox{M}\kern-.125emS}}

\title{The general relation between the weak inclusive decays of bound and free
heavy quarks\thanks{Based on the talk presented by I.M.N. at the 3$^{rd}$ 
International Conference on Hyperons, Charm and Beauty Hadrons,
Genova, June 30--July 3, 1998. This work was supported in part by the
INTAS--RFBR grants, refs. No 95--1300 and 96--155, and the RFBR grant,
ref. No 95--02--04808a}}

\author{S. Kotkovsky$^{~\rm a}$, I.M. Narodetskii$^{~\rm a}$, 
       K.A. Ter-Martirosyan
       \address{Institute for Theoretical and Experimental Physics, 117218
       Moscow, Russia}
       and 
       S. Simula
       \address{INFN, Sezione Roma III, I-00146 Roma, Italy}}
       
\begin{document}

\begin{abstract}

\noindent We derive a new parton formula for the inclusive $B$ decays and
briefly  discuss its applications to semileptonic and weak radiative decays of
the $B$-meson. 

\end{abstract}

\maketitle

\section{INTRODUCTION}

\indent The inclusive $B$ decays offer the most direct way to determine the
$CKM$ parameters $|V_{cb}|$ and $|V_{ub}|$ and the internal structure of the
$B$-meson. Both tasks complement each other: an understanding of the connection
between quark and hadron properties is a necessary prerequisite for a precise 
determination of the $CKM$ matrix.

\indent Weak $B$ decays are particular simple in the limit $m_b \to \infty$,
when the decay rate of the $B$-meson is completely determined by the decay
rate of the heavy quark itself. The account of the soft degrees of freedom 
generates important pre-asymptotic contributions due to the binding effects 
and Fermi-motion of the heavy quark inside the hadron.

\indent The leading non-perturbative correction is described by a shape  
function $F(x)$ which governs the light-cone ($LC$)  momentum distribution of
the heavy quark inside the hadron. Here $x = p^+_b / P^+_B$ is the $LC$
fraction of the heavy-quark momentum in the $B$-meson. The shape function 
arises as a result of resummation of an infinite set of leading twist
corrections in the Heavy Quark Expansion \cite{DF} and has been 
incorporated into phenomenological models of inclusive decays \cite{AEO82}. 
The impact of the Fermi motion has also been addressed in the parton models of 
Refs. \cite{JPP94}-\cite{GNST97}. The approach of these works was based on the
hypothesis of quark-hadron duality, which assumes that when a sufficient
number of exclusive hadronic decay  modes is summed up, the decay probability
into hadrons equals the one into the free  quarks. The $b$-quark was treated
as a virtual particle with mass $m_b^2 = x^2 M_B^2$, where $M_B$ is the $B$
meson mass, while the effects due to the transverse motion of the $b$--quark
were neglected. The final expressions obtained for the semileptonic branching
fractions exhibit a close analogy with deep inelastic  lepton-nucleon
scattering.

\indent The aim of this talk is to discuss the generalization \cite{KNST97} of
the work of refs. \cite{JPP94}-\cite{GNST97}. The new features are the
treatment of the $b$-quark as an on-mass-shell particle and the inclusion of
the effects of the $b$-quark transverse momenta. The main result is the
derivation of a new parton formula for the  inclusive width, which is similar
to the one derived by Bjorken  et al. \cite{BDT92} in case of infinitely heavy
$b$- and $c$-quarks:
 \begin{equation}
    \label{1.3}
    \frac{d\Gamma_B}{dq^2} = \frac{d\Gamma^{(free)}}{dq^2} ~ R(q^2),
 \end{equation}
where $d\Gamma^{(free)}/dq^2$ is the free-quark differential decay rate 
and the function $R(q^2)$ incorporates the non-perturbative effects. In Eq. 
(\ref{1.3}) $q$ is the 4-momentum of the $W$-boson. The structure of Eq.
(\ref{1.3}) suggests that in the limit $m_b \to \infty$, $m_c \to \infty$ one
has $R(q^2) = 1$, which means that the inclusive width of the $B$-meson is
the same as the inclusive width at the free quark level. At the physical
$b$-quark masses the corrections due to the bound-state factor $R(q^2)$ are
important and sensitive to the values of quark masses.

\section{A NEW PARTON APPROXIMATION}

The derivation of Eq. (\ref{1.3}) is straightforward. The modulus squared 
of the $B$-meson decay amplitude is $|{\cal M}_B|^2 = (G^2_F/2) |V_{q'b}|^2
l_{\alpha\beta} W_{\alpha\beta}$, where $l_{\alpha\beta}$ and $W_{\alpha\beta}$
are the leptonic and hadronic tensors, respectively. Following the above 
assumption the hadronic tensor $W_{\alpha\beta}$ is given through the optical
theorem by the imaginary part of the quark box diagram describing the forward
scattering amplitude:
 \begin{eqnarray}
    \label{2.3}
    W_{\alpha\beta} & = & \int\limits^1_0 \frac{dx}{x} \int d^2p_{\bot}
    w_{\alpha\beta}(p_{q'}, p_b) \cdot \nonumber\\
    & & \delta[(p_b - q)^2 - m^2_{q'}] |\psi(x,  p^2_{\bot})|^2,
 \end{eqnarray}
where 
 \begin{equation}
    w_{\alpha\beta} = w_{\alpha\beta}(p_{q'}, p_b) = \frac{1}{2} \sum_{spins}
    \bar{u}_{q'} O_{\alpha} u_b \cdot \bar{u}_b O^+_{\beta}u_{q'}
 \end{equation}
is the parton matrix element squared. The normalization of the distribution
function is given by $\int\limits^1_0 dx \int d^2p_{\bot} |\psi(x,
p^2_{\bot})|^2 = 1$. We substitute Eq. (\ref{2.3}) into the expression for
$|{\cal M}_B|^2$ and note that the contraction $l_{\alpha\beta
}w_{\alpha\beta}$ defines the modulus squared of the $b$-quark decay
amplitude: $|{\cal M}_b|^2 = (G_F^2/2) |V_{q'b}|^2 l_{\alpha\beta}
w_{\alpha\beta}$. Then, one obtains
 \begin{eqnarray}
    \label{2.5}
    |{\cal M}_B|^2 & = & |{\cal M}_b|^2 \int\limits^1_0 \frac{dx}{x} \int
    d^2p_{\bot} \cdot \nonumber \\ 
    & & \delta[(p_b - p_{q'})^2 - m_{q'}^2]| \psi(x, p^2_{\bot})|^2.
 \end{eqnarray}
We now integrate the r.h.s. of Eq. (\ref{2.5}) over $d^2p_{\bot}$, using the
$LC$ variables, to obtain Eq. (\ref{1.3}) with
 \begin{equation}
    \label{2.6}
    R(q^2) = \int\limits^{q_0^{(2)}}_{q_0^{(1)}} dq_0 ~ \omega(q_0, q^2).
 \end{equation}
The function $\omega(q_0, q^2)$ is defined as
 \begin{equation}
    \label{2.7}
    \omega(q_0, q^2) = \frac{2\pi m^2_b}{q^+} \frac{|{\bf q}|}{|{\bf \tilde
    q}|} \int\limits^{min[1,x_2]}_{x_1} dx |\psi(x ,p^{*2}_{\bot})|^2,
 \end{equation}
where $p^{*2} = m^2_b [\xi (1 - \rho - t) - \xi^2 t - 1]$, $\xi = x M_B /
q^+$, $\rho = m^2_{q'} / m^2_b$,  $t = q^2 / m_b^2$, $q^+ = q_0 + |{\bf q}|$,
and $|{\bf \tilde q}|$  is defined in the $b$-quark rest frame.  

\indent The integration limits in Eq. (\ref{2.7}) follow from the condition 
$p^{*2} \ge 0$, while the region of integration in Eq. (\ref{2.6}) is 
characterized by a quark threshold, defined through the condition 
$x_1 = min[1, x_2]$. For more details see \cite{KNST97}.

\section{SEMILEPTONIC $B$ DECAYS}

\indent The non-perturbative ingredient in Eq. (\ref{2.7}) is the $LC$ wave
function $\psi(x, p^2_{\bot})$. In what follows, we will adopt for the latter 
both a phenomenological ans\"atz suggested in \cite{MTM96} (Model $A$) and the
$LC$ wave functions corresponding to the various equal time ($ET$) wave
functions derived in quark potential models (Models $B$ to $E$). There is a
simple operational (and unitary) connection between $ET$ and $LC$ wave
functions (see, e.g., ref. \cite{C92}), which allows to convert the $ET$ wave
function into a relativistic $LC$ wave function. The latter ones have been
already used to calculate the form factors of heavy-to-heavy and
heavy-to-light exclusive transitions \cite{FF}. In this contribution we use
the $LC$ wave functions corresponding to the quark models of refs. \cite{QM}
(cases $B$ to $E$).

\indent The main difference among the various quark models relies in the
behaviour of the $ET$ wave function at high momenta. Models $B$ and $C$ 
corresponds to a soft Gaussian ans\"atz, whereas the $ET$ wave functions
corresponding to models $D$ and $E$ exhibit high momentum components generated
by the one-gluon-exchange part of the interquark potential. 

\indent We have calculated Eqs. (\ref{1.3}) and (\ref{2.6}-\ref{2.7}) for
the 
inclusive decays $B \to X_c(X_u) \ell \nu_{\ell}$ adopting the five models $A$
to $E$. In all cases the nonperturbative effects lead to a suppression with
respect to the free-decay rate. The suppression factor varies in the range
from 0.69 (case $D$) to 0.95 (case $A$) and it is sensitive both to the width
of the $x$-distribution and to the quark masses $m_b$ and $m_{q'}$. We should
stress that the $ET$ wave functions are derived from quark models containing
constituent quark masses  constrained phenomenologically by $B$- and $D$-meson
spectroscopy. Thus, for each of the quark model considered  we get a different
free-quark result. The values predicted for $\Gamma_{SL} / \Gamma_B^{(exp)}$ 
exhibit a model dependence of about $20 \%$. The average over the various 
quark model predictions yields $|V_{cb}| = (40.8 \pm 0.7_{exp} \pm 1.7_{th})
\cdot 10^{-3} \cdot \sqrt{BR_{SL}^{(exp)} / 10.43\%} \cdot \sqrt{1.57 ~ ps /
\tau_B^{(exp)}}$ (assuming the world average value $BR^{(exp)}_{SL} = (10.43
\pm 0.24) \%$) and  $|V_{ub}| = (3.53 \pm 0.44_{exp} \pm 0.14_{th}) \cdot
10^{-3}\cdot \sqrt{BR_{SL}^{(exp)} / 0.16\%} \cdot \sqrt{1.57 ~ ps /
\tau_B^{(exp)}}$. Our result for $|V_{ub}|$ is based on the preliminary ALEPH
measurement $BR(b \to u \ell\nu_{\ell}) = (0.16\pm 0.04) \%$ \cite{ALEPHW}.

\section{WEAK RADIATIVE DECAYS}

\indent The same approach can be applied to the inclusive radiative decay 
$B \to X_s \gamma$ that is one of the central decays in the rare decays 
phenomenology. The motion of the $b$-quark inside the $B$-meson leads to a
modification both of the inclusive rate and the photon energy spectrum. The
$LC$ partonic approximation for $\Gamma(B \to X_s \gamma)$ is evaluated using
the magnetic photon penguin. The result is $\Gamma(B \to X_s \gamma)=
\Gamma^{(free)}(b \to s \gamma) ~ R_{\gamma}$, where
 \begin{equation}
    \label{ratio}
    R_{\gamma} = 2m_b \pi \int\limits_0^{M_B/2} dq_0 \int\limits_{x_{min}}^1
    dx f(x, p^{*2}_{\bot}),
 \end{equation} 
with $p_{\bot}^{*2} = m_b^2(x / x_{min} - 1)$, and $x_{min} = 2q_0 / M_B$, the
mass of the strange quark being neglected. In Table 1 the coefficients
$R_{\gamma}$ obtained using the above ans\"atz for the $LC$ wave function are
collected. We also report the branching fractions $BR_{\gamma} = BR(B \to
X_s \gamma)$ calculated using the quark models $A$ to $E$, showing the 
sensitivity of the theoretical predictions to the different choices of the 
$ET$ wave functions and constituent quark masses entering the calculation. We
use $|C^{(0)eff}_{7_{\gamma}}| \approx 0.3$ and $1 / \alpha_{EM} = 137$.
Averaging over the models one gets $BR_{\gamma} = (3.3^{+0.7}_{-0.9}) \cdot
10^{-4}$ that agrees with the very recent preliminary update from CLEO,
$BR_{\gamma} = (2.50 \pm 0.47 \pm 0.39) \cdot 10^{-4}$ \cite{CLEO}, and the
ALEPH measurement, $BR_\gamma = (3.11 \pm 0.80 \pm 0.72) \cdot 10^{-4}$
\cite{ALEPH}, within one standard deviation.

\section*{ACKNOWLEDGMENTS}

\noindent One of the authors (I.M.N.) would like to thank Carlo Caso and Calvin
Kalman for organizing an excellent Conference with a stimulating scientific
program. 

%%%%%%%%%%%%%%%%%%%%%%%%%%%%%%%%%%%%%%%%%%%%%%%%%%%%%%%%%%%%%%%%%%%%%%%

\begin{table}[t]
\caption{The ratio $R_{\gamma}$ (Eq. (\protect\ref{ratio})) and the branching
fraction $BR_{\gamma}$ (in units $10^{-4}$) calculated within the quark models
$A$ to $E$.}
\begin{center}
\begin{tabular*}{70mm}{@{}l@{\extracolsep{\fill}}rrrrr}  
\hline
  Model & A & B & C & D & E  \\ \hline 
  $R_{\gamma}$ & $0.96$ & $0.94$ & $0.88$ & $0.88$ & $0.88$ \\ 
  $BR_{\gamma}$ & $2.4$ & $2.8$ & $3.8$ & $4.0$ & $3.5$ \\ \hline
\end{tabular*}
\end{center}
\end{table}

%%%%%%%%%%%%%%%%%%%%%%%%%%%%%%%%%%%%%%%%%%%%%%%%%%%%%%%%%%%%%%%%%%%%%%%%


\begin{thebibliography}{99}

\bibitem{DF} I. Bigi et al., Int. J. Nucl. Phys. A9 (1994) 2467; M. Neubert,
 Phys. Rev. D49 (1994) 3392
\bibitem{AEO82} G. Altarelli et al., Nucl. Phys. B208 (1982) 519
\bibitem{JPP94} C.H. Jin et al., Phys. Lett. B329 (1994) 364
\bibitem{MTM96} V.L. Morgunov and K.A. Ter-Martirosyan, Phys. of Atom. Nucl. 59
 (1996) 1221
\bibitem{GNST97} I.L. Grach et al., Nucl. Phys. B502 (1997) 227
\bibitem{BDT92} J. Bjorken et al., Nucl. Phys. B371 (1992) 111
\bibitem{KNST97} S. Kotkovsky, Pis'ma v ZhETP 65 (1997) 734; S. Kotkovsky
et al., hep-ph/9712543.
\bibitem{C92} F. Coester, Prog. Part. Nucl. Phys. 29 (1992) 1
\bibitem{FF} See e.g. N.B. Demchuk et al., Phys. Lett. B409 (1997) 2152 and
 references therein quoted
\bibitem{QM} V.O. Galkin et al., Yad. Fiz. 55 (1992) 2175 (case B); I.M.
 Narodetskii et al., J. Phys., G18 (1992) 2175 (case C); D. Scora, N.Isgur, 
 Phys. Rev. D52 (1995) 2783 (case D); S. Godfrey, N.Isgur, Phys. Rev., D32
 (1985) 185 (case E)
\bibitem{ALEPHW} ALEPH collaboration: contribution PA05-059 to the XXVIII
 Int. Conf. on High Energy Physics, Warsaw, Poland, 1996
\bibitem{CLEO} S. Glenn, talk presented at the Meeting of the APS, Columbus,
 Ohio, 18-21 March 1998
\bibitem{ALEPH} R. Barate et al., CERN-EP/98-044 

\end{thebibliography}
\end{document}